# Electrically Driven Light Emission from Individual CdSe Nanowires

Yong-Joo Doh,[†,∥] Kristin N. Maher,[†] Lian Ouyang,[†] Chun L. Yu,[†] Hongkun Park[†,‡,*]

and Jiwoong Park,[§,¶,*]

[†]Department of Chemistry and Chemical Biology, Harvard University, Cambridge, MA 02138, USA, [∥]National CRI Center for Semiconductor Nanorods and Department of Materials Science and Engineering, POSTECH, Pohang 790-784, Korea, [‡]Department of Physics, Harvard University, Cambridge, MA 02138, USA, [§]The Rowland Institute at Harvard, Cambridge, MA 02142, USA, [¶]Department of Chemistry and Chemical Biology, Cornell University, Ithaca, NY 14853, USA

*Authors to whom correspondence should be addressed. E-mail: (HP) Hongkun_Park@harvard.edu and (JP) jp275@cornell.edu and

ABSTRACT

We report electroluminescence (EL) measurements carried out on three-terminal devices incorporating individual *n*-type CdSe nanowires. Simultaneous optical and electrical measurements reveal that EL occurs near the contact between the nanowire and a positively biased electrode or drain. The surface potential profile, obtained by using Kelvin probe microscopy, shows an abrupt potential drop near the position of the EL spot, while the band profile obtained from scanning photocurrent microscopy indicates the existence of an *n*-type Schottky barrier at the interface. These observations indicate that light emission occurs through a hole leakage or an inelastic scattering induced by the rapid potential drop at the nanowire-electrode interface.

Semiconductor nanowires (NWs) have emerged as promising materials for integrating electronics with photonics on a single nanoscale platform.[1,2] Significant advances have been made toward highly controlled synthesis of single-component and heterostructure NWs, opening rational routes for novel devices that utilize their size-dependent optical and electrical properties.[1-3] A variety of optoelectronic devices have already been demonstrated using these NWs and NW heterostructures, such as photo-detectors,[4] lasers,[5] and light-emitting devices (LEDs).[6,7] These devices also showed the promise in controlling the optoelectronic response at the single-quantum level.[8]

Typically NW LEDs have adopted a *p-n* junction structure[6,7] where the electrons and holes are injected into the junction to recombine radiatively. It has also been reported, however, that electroluminescence (EL) from single-component nanocrystals[9] or NWs[10] can also occur through the inelastic scattering by tunneling electrons[9] or the impact excitation process by hot carriers.[10] Similar EL behaviors have also been studied using carbon nanotube (CNT) devices.[11-16] In order to gain insights into the light emitting mechanism in the three-terminal devices incorporating individual nanostructures, it is essential to carry out complementary study of the electronic transport and the optical measurements that allow spatially resolved probing of device properties, especially the metal-nanostructure junction.

In this letter, we report the bias-dependent EL measurements from the individual *n*-type CdSe NWs, a prototypical direct band-gap semiconductor with a band gap of $E_{gap}$ = 1.70 eV. Spatially and energy resolved information regarding the EL signal is obtained as a function of source-drain bias and gate voltage. The local electric potential of NW devices and energy band profiles near the electrode contacts are determined by scanning photocurrent microscopy (SPCM)[17,18] and Kelvin probe microscopy (KPM).[19] Spatially-localized EL spot is observed near a positively-biased contact or drain where a rapid potential drop occurs. These observations provide an important insight into the mechanisms for light emission in these devices.

CdSe nanowires were synthesized using a solution-liquid-solid approach with Bi nanoparticles as catalysts.[20] After depositing nanowires onto a degenerately doped silicon wafer covered with a 300 nm-thick oxide layer, standard electron-beam lithography was used to define source and drain electrodes. After resist development, the exposed nanowire surface was treated with oxygen plasma and a chemical etchant ($H_2O$:HCl:$HNO_3$ = 12:3:1 by volume), followed by thermal evaporation of In (120 nm) onto the liquid-nitrogen-cooled sample. After lift-off, a passivating layer of $SiO_2$ or $Si_3N_4$ (50 nm) was deposited using electron-beam evaporation. The electrical conductance of a device without a passivation layer was much smaller than that of one with a passivation layer, most likely due to the presence of adsorbed oxygen around the nanowire.[21] A scanning electron microscope image of a resulting device is shown in Fig. 1a. The diameter of CdSe NWs ranged from 45 to 65 nm and the source-drain spacing ranged from 2 to 6 μm. Electrical conductance measurements without light illumination at room temperature show that NW devices exhibit *n*-type conductance (see Fig. 2c) with an average electron density $n_s$ = ~5 x $10^{18}$ $cm^{-3}$ and an electron mobility $\mu$ = (0.5 - 8) $cm^2$/Vs.

Single-nanowire EL and photoluminescence (PL) measurements were carried out using an optical microscope outfitted with an intensified charge-coupled device (CCD) camera and a spectrometer. All PL and EL spectra were divided by a system response curve obtained using a calibrated light source to correct for unequal sensitivity to different wavelengths. For SPCM measurements, a diffraction-limited laser spot with a wavelength of 532 nm was scanned over the device, while the resulting photocurrent $I_{ph}$ and the reflected light intensity were simultaneously recorded as a function of the laser spot position. In KPM measurements, an atomic force microscope with a conductive tip was used to measure the contact potential difference between the tip and the sample while the topography of the nanowire device was simultaneously measured. All these measurements were carried out in air at room temperature. Further details of the SPCM and KPM measurements can be found in earlier reports.[17, 22]

Figure 1a shows a scanning electron microscope (SEM) image of a typical device. Applied with a high bias of *V* = 7 V, spatially localized EL emission is observed near the positively-biased contact, as shown in Fig. 1b. Under the sign reversal of the bias with *V* = -7 V, the EL spot occurs near the counter electrode (see Fig. 1c). The corresponding current-voltage (*I-V*) characteristic curve is displayed in Fig. 1d, overlaid with the integrated EL intensity which was measured simultaneously. It is clearly seen that light emission occurs above a threshold bias $V_{th} \sim \pm 4$ V at the the positively biased contact. More than 13 devices have shown similar behavior of the bias-dependent EL.

Above $V_{th}$, the EL intensity increases quadratically with the current, as shown in the inset of Fig. 1d. This observation indicates that the quantum efficiency $\eta$ for the light emission increases linearly on the current, resulting in $\eta = (1 - 5) \times 10^{-6}$ at $I = 1$ μA. This $\eta$ value is close to that observed from the InP nanowire p-n junctions[6] and an order of magnitude larger than that from a single CNT field effect transistor.[13] The EL spectra with different $V$, shown in Fig. 1e, exhibit a relatively broad emission peak centered at the photon energy of ~1.66 eV. This value is close to a band gap ($E_{gap} = 1.70$ eV) of the CdSe NW obtained from the photoluminescence (PL) spectrum of a single NW (see the inset). The full width at half maximum is about 0.51 (0.09) eV for the EL (PL) spectrum at room temperature.

To understand the physical mechanism of the bias-dependent EL near the contact, we investigated the electronic band structure near the metal-NW junction by SPCM. Figure 2b shows a spatially resolved photocurrent image taken at $V = V_g = 0$ V obtained from a CdSe NW device. The most notable feature in this image is the presence of local steady-state photocurrent ($I_{ph}$) spots near both electrode contacts. The position and polarity of these photocurrent spots indicate that a local upward band bending occurs near the interface between the CdSe NW and the metal electrode, implying the formation of n-type Schottky barriers at the contacts.[17, 18] By comparing $I_{ph}(V_g)$ curve with dark-state $G(V_g)$ curve in Fig. 2c,[17] the Schottky barrier heights are estimated to be be $\phi_R = 0.52$ eV and $\phi_L = 0.74$ eV for the right and left contact, respectively.[23]

The KPM measurements, shown in Fig. 3a, allows the spatial profiling of the surface potential of the NW device with different $V$. Specifically, Fig. 3a displays the line profile of the Kelvin potential ($V_{Kelvin}$) along the NW, obtained from the scanning KPM. With increasing $V$, an asymmetric potential profile develops with a major potential drop occurring always near a positively biased contact. The position of this rapid potential drop coincides with the location of the EL spot. More than 7 devices showed similar behavior of $V_{Kelvin}$ profile.

By comparing the local $V_{Kelvin}$ value at each electrode with the one at the middle position of the NW, we can determine the regional potential drops of $\Delta V_L$ and $\Delta V_R$ at the left- and right-segments. Figure 3b shows $\Delta V_L$ and $\Delta V_R$ as a function of $V$, overlaid with the EL intensity plot. It can be clearly seen that the development of the asymmetric potential profile and the light emission occurs at a similar bias range.

The electrically driven light emission from a single-component nanostructures can be explained by several different scenarios: these include a bipolar EL mechanism where the electrons and holes are injected simultaneously into the nanostructures,[11, 12, 16] an unipolar impact excitation process by hot carriers,[9, 10, 14] and a thermal light emission due to Joule heating.[15] Among these scenarios, the thermal heating can be ruled out as a source of the observed EL by two experimental observations. First, the negative differential conductance, which is indicative of significant Joule heating, is not observed in our *I-V* curve. Second, our observed EL spot is located near one of the contacts rather than the center of the NW and changes its position depending on bias polarity even with the same electrical power of the NW device, again incompatible with the Joule heating mechanism.[16]

In order for the light emission to be explained by the bipolar EL mechanism, on the other hand, the device should exhibit ambipolar characteristics so that the injection of electrons and holes can occur at the same time.[11, 12] The exciton recombination rate should then be limited by the minority carrier current, and the EL intensity should be strongest near the ambipolar region of $V_g$ rather than *n*- or *p*-type region.[11, 12] As shown in Fig. 2c, however, our CdSe NW devices always exhibit *n*-type characteristics. Moreover the $V_g$-dependent EL intensity plot in Fig. 4a indicates that the light emission is enhanced monotonically with $V_g$, contrary to the previous results in the ambipolar CNT.[11]

Recently it was suggested that the bipolar EL mechanism is also possible even in the case of exclusive unipolar devices.[16] Under high current operation, the majority carriers (electrons, here) can be backscattered and accumulated near a drain contact to promote the leakage of the minority carriers (holes) from the drain electrode.[16] Then the EL emission can occur through the radiative electron-hole recombination near the contact, consistent with our observations in Fig. 1b and c. In this scenario, the poor hole mobility[21, 24] of the *n*-type NW explains the immobile EL spot in our CdSe NW devices. Since the tunneling rate of minority carriers would be proportional to the number of the accumulated space charges, the EL intensity would increase with the total current, which also gives a qualitative explanation for the *n*-type behavior of the $V_g$-dependent EL intensity data in Fig. 4a. Furthermore, the abrupt potential drop near the drain can be due to the annihilation of electrons and holes at the emission spot, as was invoked in ambipolar CNT devices.[25, 26]

Another possible explanation for our observed EL is based on an impact excitation process[27] under unipolar electrical transport.[14] Since the abrupt potential drop near the drain or positively-biased contact can generate a strong electric field, the electrons are accelerated to scatter inelastically and produce electron-hole pairs, which recombine radiatively. The EL intensity ($I_{EL}$), which would be proportional to the impact excitation rate within this model, can be represented by $I_{EL} \sim \exp(-F_{th}/F_{loc})$ where $F_{th}$ is a threshold electric field and $F_{loc}$ is a local electric field in the light emitting region.[14] To a first approximation, $F_{th}$ is given by 1.5 $E_{gap}/e\lambda_{ph}$ ($\lambda_{ph}$ is the optical phonon scattering length),[27] and $F_{loc}$ is approximately $V/w$ where $w$ is the length of the NW segment emitting the light. Using the values of $E_{gap}$ = 1.70 eV and $\lambda_{ph}$ ~ 5 nm that correspond to our CdSe NWs,[28] $F_{th}$ is then estimated to be ~ 5.1 MV/cm, similar to that of bulk GaAs[28] but 10 times larger than that of a carbon nanotube.[14] Figure 4b shows semi-log plot of the bias-dependent EL intensity with a fit to the above impact excitation model for three different devices. From the fit the length of the NW segment for an active EL zone is estimated to be $w$ = 50 – 75 nm. In this picture, the relatively broad EL spectrum shown in Fig. 1e can be naturally explained by the short time scale associated with the inelastic scattering process.[13] The low EL efficiency is due to the low efficiency of the inelastic scattering process.

Although the data presented in this letter enables detailed examination of various theoretical models and allows the discrimination between them, they are nevertheless not enough to distinguish between the bipolar EL mechanism and the unipolar impact excitation process. Clearly, further theoretical and experimental investigation would be needed to make a conclusive remark regarding these two scenarios. Under the interplay of the carrier transport and the radiative recombination process, however, it is evident that the strong electric field localized near a drain contact plays an important role in the light emission of *single-component* NW devices, which is a common feature in both mechanisms. We suggest that the band-edge emission from a single-component NW would be a general behavior in the other semiconductor NWs with direct band gaps.

**Acknowledgments**: We thank Y. Ahn, D. Yu, Q. Gu, G.-T. Kim, and G.-C. Yi for useful discussions. YD was partially supported by the National Creative Research Initiative Project (R16-2004-004-01001-0) of the Korea Science and Engineering Foundations (KOSEF).

FIGURE CAPTIONS

**Figure 1**. (a) Scanning electron microscope image of a CdSe NW device, **D1**. (b) EL image overlaid with the reflection image of a device with a bias voltage of $V = 7$ V applied to the top electrode. Spatial resolution is 130 nm, limited by CCD pixel size. (c) As for (b), with $V = -7$ V. (d) Current ($I$) and simultaneously measured EL intensity plotted against bias voltage with gate voltage $V_g = 0$ V. Inset: log-log plot of EL intensity versus current from four different devices **D1** (triangle), **D3** (square), **D4** (open circle) and **D5** (closed circle). (e) EL spectra for **D1** obtained at different bias voltages $V = 4, 5, 6, 7,$ and 8 V from bottom to top. Inset: PL spectrum of a single CdSe NW using a diode laser with a wavelength of 405 nm for continuous excitation. The peak is centered at the semiconductor band gap $E_{gap} = 1.7$ eV.

**Figure 2.** (a) Reflection image of **D2**. The inset depicts our measurement scheme for SPCM and a definition of the direction of positive photocurrent. (b) Corresponding scanning photocurrent image taken simultaneously at $V = V_g = 0$ V. Red (blue) spot corresponds to a positive (negative) photocurrent of 141 pA (-170 pA) at maximum (minimum). The wavelength of the laser was 532 nm with an intensity of 430 W/cm$^2$. The dashed line indicates the location of the NW. (c) $V_g$ dependence of the conductance ($G$) in the dark state and photocurrent $I_{ph}$ with local illumination at the left (blue) and right (red) contact in the linear response regime for **D2**. Inset: energy band diagram with $V = 0$ V, where $\phi_L = 0.74$ eV ($\phi_R = 0.52$ eV) is the barrier height at the left (right) contact and $\Delta\phi = 0.18$ eV is the energy difference between the Fermi level and the bottom of the conduction band.

**Figure 3.** (a) KPM measurements of the local potential profile along the length of the NW from **D3**. The bias voltage is varied from -4.5 V (bottom) to 4.5 V (top) with an increment of 0.5 V. The left electrode is grounded with bias applied to the right electrode. The dashed lines indicate the electrode

boundaries. The difference between measured KPM potential and the applied bias voltage is due to an electrostatic coupling between the tip and the whole sample.[8] Inset: AFM topography image (top) and KPM image with $V = +4$ V (middle) and -4 V (bottom) for the same device. The lift height for the KPM scan is 100 nm. Scale bar is 1 µm. (b) Regional potential drops of $\Delta V_L$ and $\Delta V_R$ corresponding to left (closed circle) and right (open circle) NW segments, respectively, plotted with $V$. The corresponding EL intensity (solid line) is overlaid. $V_c$ for the asymmetric potential profile and $V_{th}$ for the light emission are indicated by arrows, respectively.

**Figure 4.** (a) $V_g$ dependence of EL intensity (blue) and current (black) with $V = 8$ V for **D5**. (b) Bias $V$ dependence of EL intensity for **D1** (square), **D3** (triangle), and **D4** (circle). Each solid line is a fit to the impact ionization model, as explained in the text.

Doh et al., Fig. 1

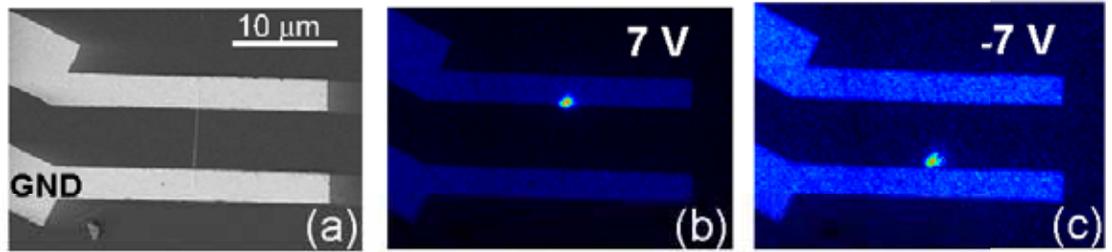
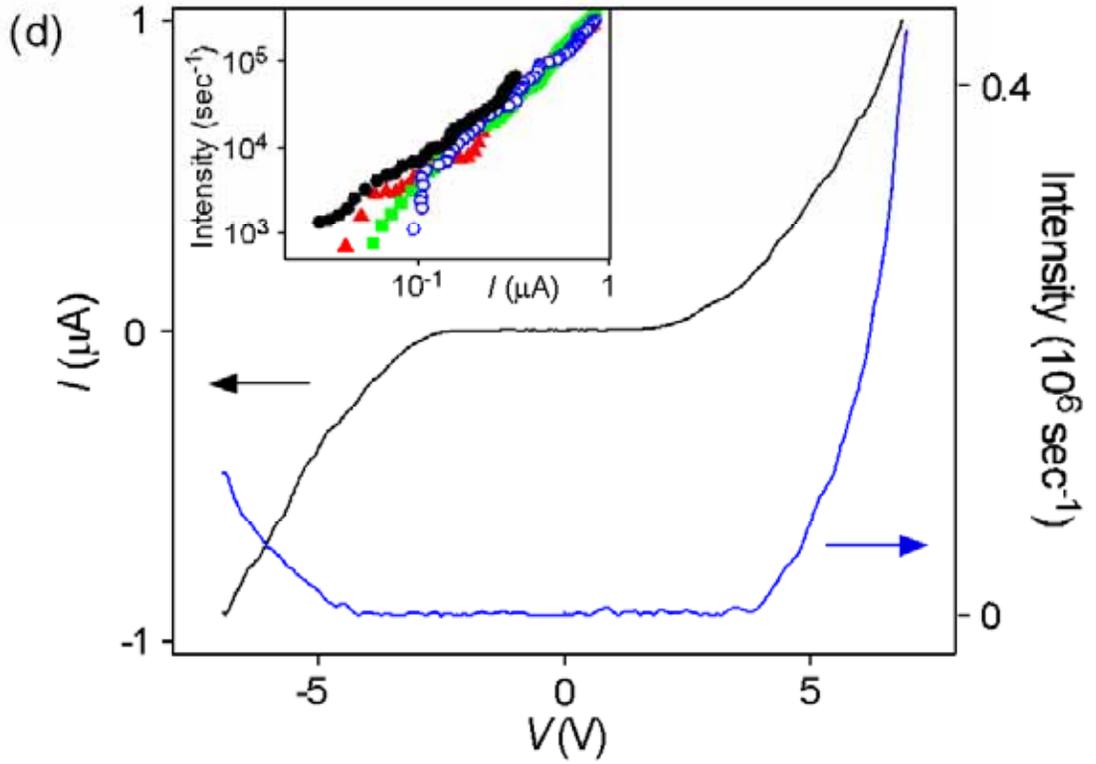
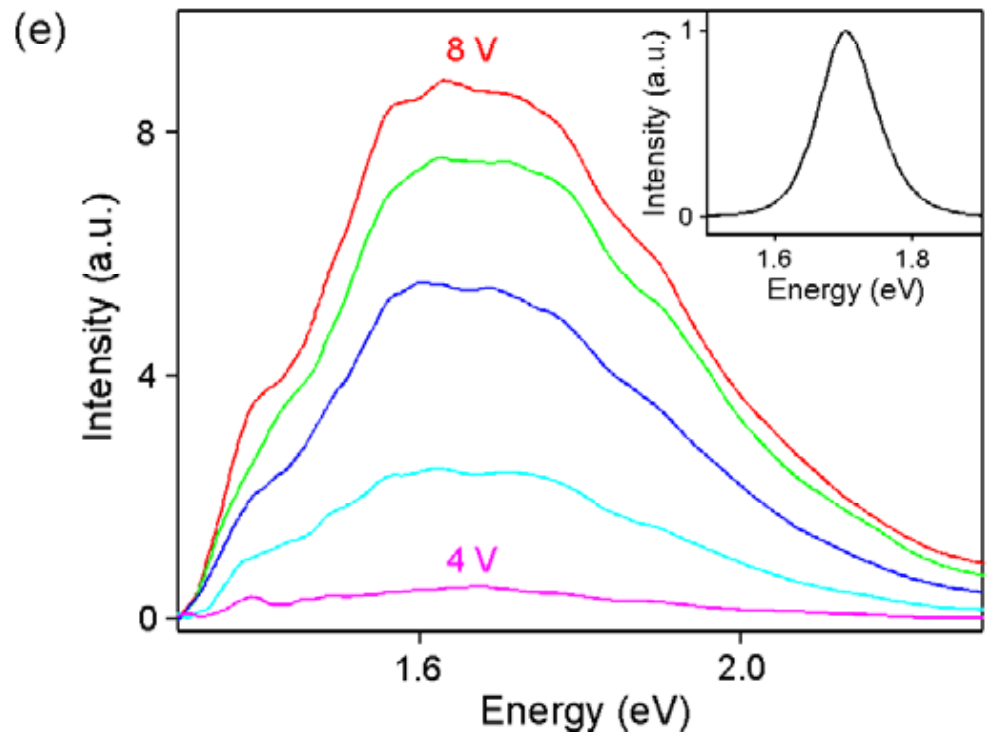

Doh et al., Fig. 2

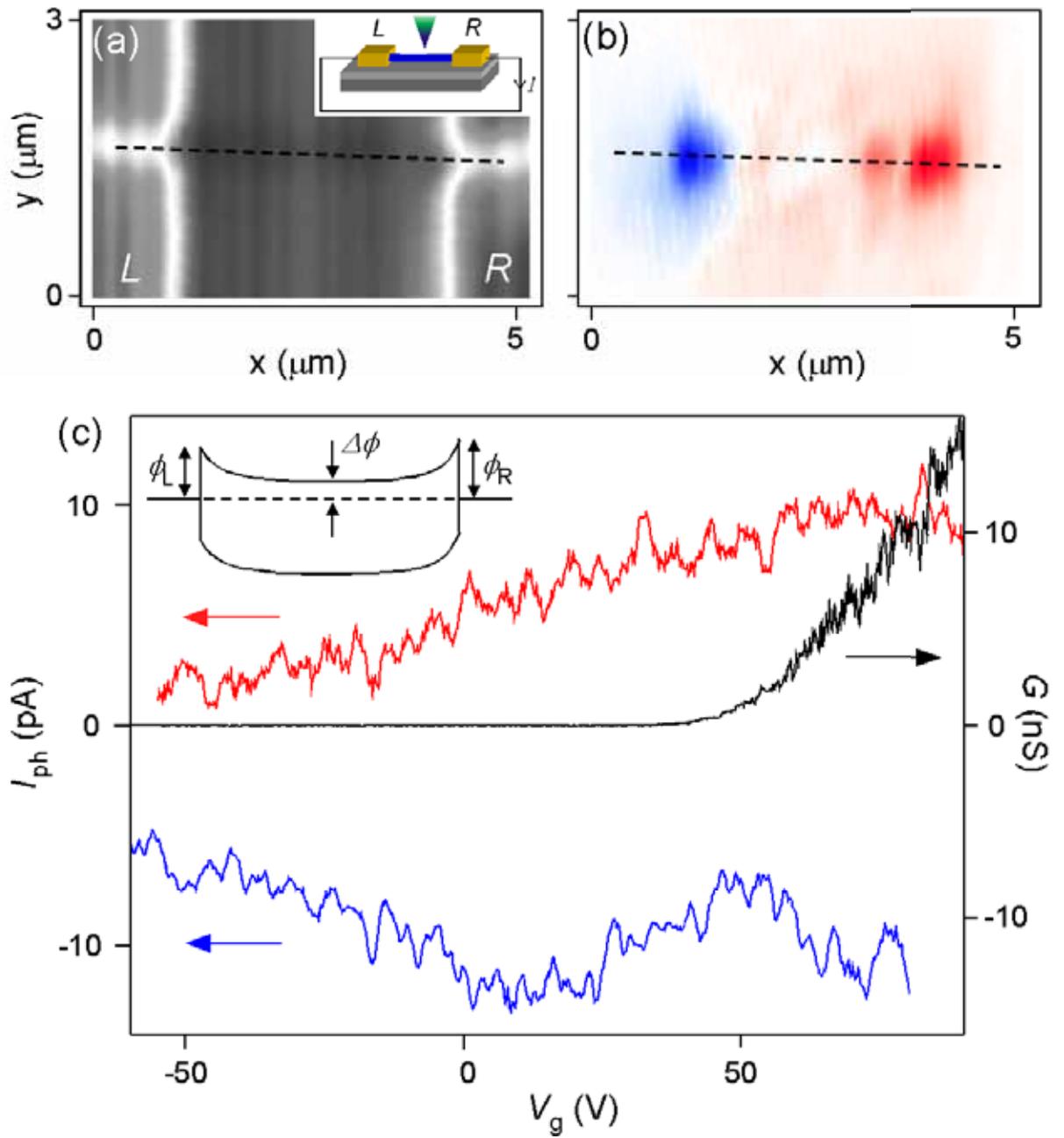

Doh et al., Fig. 3

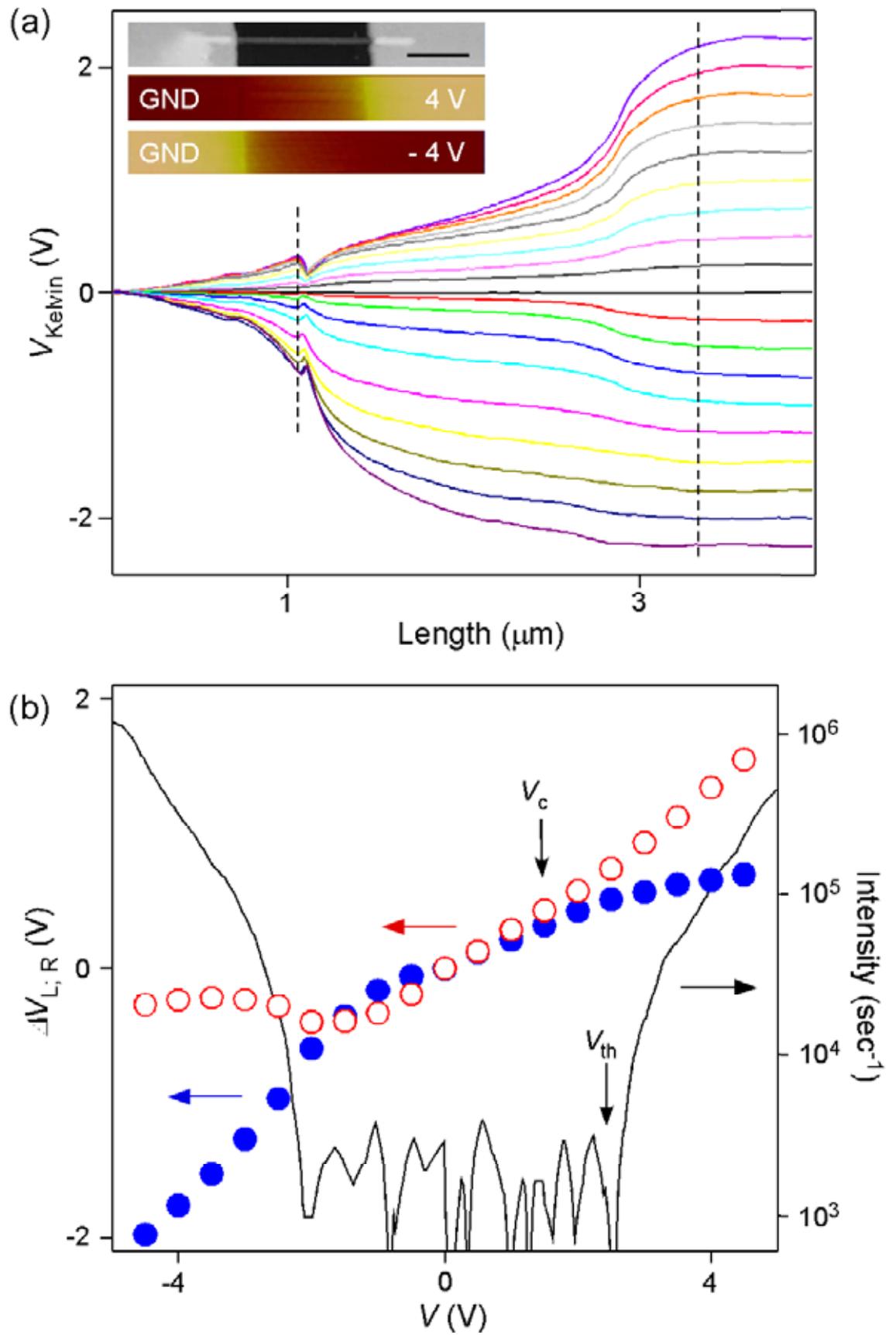

Doh et al., Fig. 4

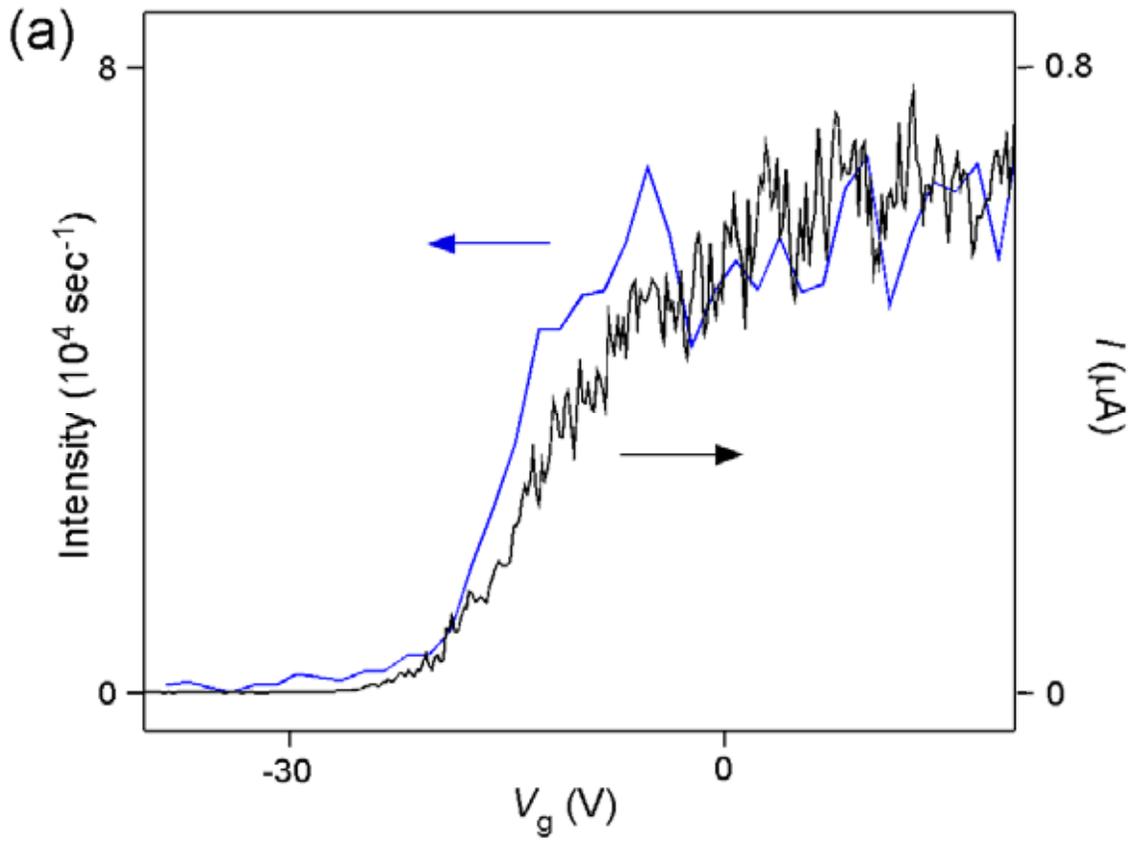

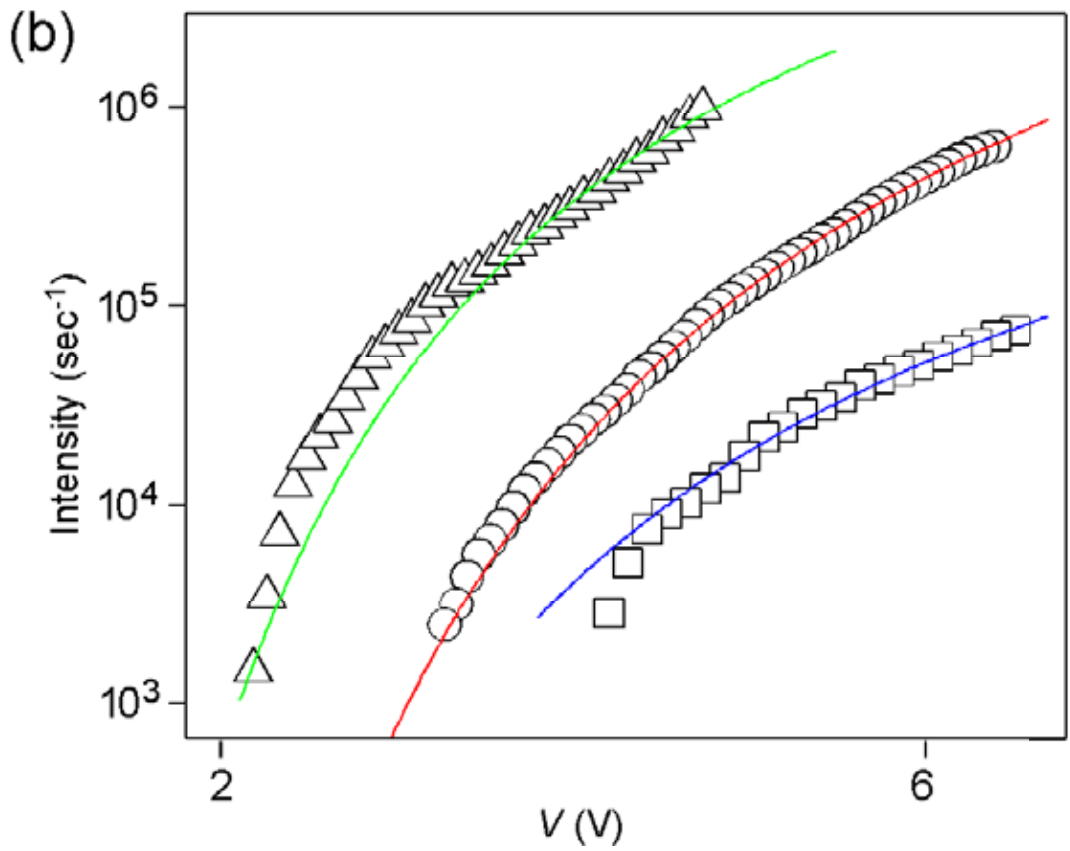

16